\documentclass[jgrga]{AGUTeX}

\usepackage{graphicx}
\setkeys{Gin}{draft=false}
\authorrunninghead{COHEN ET AL.}

\titlerunninghead{Interchange Reconnection in the Corona}

\begin{document}

%
%

\title{Numerical Simulation of the May 12, 1997 CME Event - the Role of Magnetic 
Reconnection}

%
%



\authors{O. Cohen, \altaffilmark{1} G. D. R. Attrill, \altaffilmark{1}
N. A. Schwadron, \altaffilmark{2} N. U. Crooker, \altaffilmark{2}
M. J. Owens, \altaffilmark{3} C. Downs, \altaffilmark{4} 
and T. I. Gombosi, \altaffilmark{5}}

\altaffiltext{1}{Harvard-Smithsonian Center for Astrophysics, 60 Garden St., Cambridge, MA 02138, USA.}

\altaffiltext{2}{Center for Space Physics, Boston University, 725 Commonwealth Ave., Boston, MA 02215, USA.}

\altaffiltext{3}{Space Environment Physics Group, Department of Meteorology, University of Reading, RG6 6BB, UK}

\altaffiltext{4}{Institute for Astronomy, University of Hawaii, at Manoa, 2680 Woodlawn Dr., Honolulu, HI 96822, USA}

\altaffiltext{5}{Center for Space Environment Modeling, University of Michigan, 2455 Hayward, Ann Arbor, Michigan, USA.}

%
%


\begin{abstract}

We perform a numerical study of the evolution of a Coronal Mass Ejection (CME) and its interaction with the coronal 
magnetic field based on the May 12, 1997, CME event using a global MagnetoHydroDynamic (MHD) model for the solar corona. 
The ambient solar wind steady-state solution is driven by photospheric magnetic field data, while the solar eruption 
is obtained by superimposing an unstable flux rope onto the steady-state solution. During the initial stage of CME 
expansion, the core flux rope reconnects with the neighboring field, which facilitates lateral expansion of the CME 
footprint in the low corona. The flux rope field also reconnects with the oppositely orientated overlying magnetic 
field in the manner of the breakout model. During this stage of the eruption, the simulated CME rotates counter-clockwise 
to achieve an orientation that is in agreement with the interplanetary flux rope observed at 1 AU. A significant component 
of the CME that expands into interplanetary space comprises one of the side lobes created mainly as a result of 
reconnection with the overlying field.  Within 3 hours, reconnection effectively modifies the CME connectivity from the 
initial condition where both footpoints are rooted in the active region to a situation where one footpoint is displaced 
into the quiet Sun, at a significant distance ($\approx 1R_\odot$) from the original source region. The expansion and rotation due 
to interaction with the overlying magnetic field stops when the CME reaches the outer edge of the helmet streamer belt, 
where the field is organized on a global scale. The simulation thus offers a new view of the role reconnection plays in 
rotating a CME flux rope and transporting its footpoints while preserving its core structure.

\end{abstract}

\begin{article}


\section{INTRODUCTION}
\label{sec:Intro}

In recent decades, the concept of magnetic reconnection, in which two oppositely orientated magnetic field lines are 
cut and re-assembled, has become more accepted as an important process in solar coronal dynamics. In particular, 
magnetic reconnection seems to dominate the initiation, evolution, and propagation of Coronal Mass Ejections (CMEs), 
as the magnetic flux carried by the CME interacts with the pre-existing field both in the low corona and throughout 
the heliosphere. In recent years, a number of CME initiation models have attempted to resolve the observed features 
of a solar eruption. Many CME initiation models drive the eruption by imposing photospheric shear motion, flux diffusion, and 
flux cancellation to the footpoints of the pre-existing magnetic field (see \cite[e.g.][]{forbes95,amari00,linker03} 
and the review by \cite{forbes06} and references therein). In other words, the eruption is obtained through 
modification of the boundary conditions, transforming the initial potential magnetic field into a non-potential, 
twisted magnetic field, which has the required free energy stored in it. This non-potential state introduces currents 
so that the Lorentz force overcomes the downward magnetic tension and gravity; eventually, the CME erupts when the 
force balance breaks down. In this class of CME initiation model, which we refer to as the ``driven'' model, magnetic 
reconnection occurs below the CME flux rope in the vicinity of the magnetic field footpoints.

Another type of CME initiation model is the breakout model \citep{antiochos99,lynch08}, in which a 
bipolar region (representing the source active region of the CME) is embedded in a background dipole field of opposite 
polarity. The eruption in this case is driven by shearing the footpoints of the local bipolar region around its neutral 
line.  As a result, the local bipolar field inflates and begins to reconnect with the oppositely orientated overlying 
field. This reconnection transfers overlying field to the side lobes, until the CME erupts through the weakened overlying 
field. The location of magnetic reconnection that drives this type of CME initiation model occurs above the CME flux rope, and 
there is no opening or disconnection of flux during the process.  There are two main problems with most of the current 
CME initiation models. First, these models use idealized magnetic configurations, which only mimic the realistic coronal 
field, along with idealized footpoint motions. Second, almost all models neglect the non-potential, background solar 
wind solution into which the CME is erupting \citep{manchester04,lugaz07,cohen08a}. Taking into account a more realistic 
coronal environment might significantly affect the propagation of the CME due to the interaction with the more complex 
ambient field. This three-dimensional interaction can facilitate the lateral expansion of the outer shell of the CME via 
magnetic reconnection during the eruption \citep[e.g.][]{manoharan96,pick98,pohjolainen01}.

One of the observed signatures associated with CMEs in the low corona are the so-called coronal ``waves''. Since their discovery 
in 1996 \citep{dere97,moses97,thompson98}, the physical nature of EIT coronal waves, which are 
strongly linked to CMEs \citep{biesecker02} has been under debate. In particular, there is an argument whether the observed 
coronal ``wave'' is a MagnetoHydroDynamic (MHD) wave, or whether it corresponds to the actual footprint of the CME (for a complete 
discussion on this debate see \cite{cohen09}, as well as \cite{OfmanThompson02,ofman07,Schmidtofman10} and references therein). In one 
non-wave model, the latter requires the flux rope to reconnect with the surrounding coronal field, facilitating the lateral 
expansion \citep{attrill07a}. This paper demonstrates that the concept can occur in conjunction with the breakout model where the original flux rope 
reconnects with the overlying field. The main difference between the two is essentially the location of the reconnection point. 
In the breakout model, the polarity of the CME flux rope is opposite to that of the large-scale overlying field, so that the 
reconnection occurs at the top of the flux rope. In the model by \cite{attrill07a}, which we can call the ``stepping 
reconnection model'', reconnection occurs whenever the core flux rope meets a neighboring loop with opposite polarity and the 
reconnection point is located to the side of the expanding core flux rope.  We demonstrate that the stepping reconnection model can 
theoretically occur within the central part of the breakout model topology, facilitating the expansion of the core flux rope, 
prior to interaction with the larger-scale overlying field. This concept is discussed in Section~\ref{sec:Discussion}.

In a three-dimensional MHD simulation of a recent CME event (February 13, 2009), \cite{cohen09} have shown that the core flux 
rope indeed reconnects with the surrounding field in the same manner as described by \cite{attrill07a}. They also showed that 
the bright front constituting the diffuse EIT coronal waves is composed of both a piston-driven MHD wave as well as non-wave 
components, which are coupled as long as the CME continues to expand laterally. The lateral expansion essentially stops when the 
field topology no longer allows the magnetic field to reconnect, and the magnetic overpressure of the flux rope relative to the 
surroundings no longer drives a strong lateral expansion. From this point, which can be at the considerable distance of 
$\approx 1R_\odot$ from the source region \citep{attrill09,cohen09}, only the MHD wave component continues to exist as a 
freely propagating wave.

The May 1997 CME event is a SHINE campaign event and it has been chosen due to the fact that it was an isolated halo 
CME event occurred during solar minimum, which supposedly makes it a ``simple'' event, and due to the relatively extensive observational 
data of the event. Several attempts to simulate the global evolution of a CME event in the corona and in the heliosphere have been 
done in the past decade. For example, \cite{manchester04,fangibson07,riley08,vanderholst09} (and references therein) have simulated 
CME eruptions driven by different methods into either a background corona in hydrostatic equilibrium or a more physical solar wind 
background. However, even the steady state solar wind solution in these simulations was based on an idealized dipole configuration 
for the solar magnetic field. The particular May 1997 event has been simulated by \cite{Odstrcil04,Odstrcil05,shen07,Wu07}. All 
these simulations however, were focused on the propagation of the CME to the Earth, and they were driven by kinematic propagation 
of the MHD parameters to the inner boundary of the simulation domain, which has been set to be beyond the Alfv\'enic point. 

In this work, we use a model for the solar corona which is driven by high-resolution magnetogram data and provides a more realistic 
coronal magnetic field. The model also simulates a CME that propagates through a steady-state, non-potential MHD solution for the 
solar corona and the solar wind. The advantage of such a model is that it enables us to study the complex interaction of the CME 
with the realistic ambient field. The main limitation of the model is that the CME is a fully-formed flux rope that is ``injected'' 
into the steady-state solution with its observed parameters. While reconnection between the closed loops of the flux rope and the 
surrounding magnetic field is a primary focus of this paper, any reconnection between closed loops beneath the CME that generates 
flux rope coils, as occurs in many CME initiation models, is not part of our simulation. Another difference between this model and 
the CME initiation models described previously is that here we are interested in the interaction above the surface with fixed 
photospheric boundary conditions, while in the other models, modification of the boundary conditions is the main driver for the 
simulation. We present a high-resolution numerical simulation of the complex interaction between the erupting CME and the ambient 
coronal field (up to $24R_\odot$) based on the May 12, 1997, CME event. The goal of this work is to better understand the 
three-dimensional interaction of the CME with the ambient flux.  

The structure of this paper is as follows. We describe the numerical simulation in Section~\ref{sec:Model} and present the results 
in Section~\ref{sec:Results}. We discuss the consequences and implications of the simulation results in Section~\ref{sec:Discussion} 
and conclude our findings in Section~\ref{sec:Conclusions}.


\section{Numerical Model}
\label{sec:Model}

\subsection{Ambient Solar Wind and CME Initiation}

For our study, we use the Solar Corona (SC) module \citep{cohen07} of the Space Weather Modeling Framework 
(SWMF) \citep{toth05}, which is based on the BATS-R-US global MHD code \citep{powell99}. 
The steady-state solar wind solution is obtained in a non-polytropic manner \citep{roussev03b,cohen07,cohen08b} 
and it is constrained by the empirical Wang-Sheeley-Arge (WSA) model \citep{wang90,argepizzo00}. 
We use a potential magnetic field \citep{altschuler69} to prescribe the initial condition for the magnetic 
field using high-resolution MDI magnetograms (http://sun.stanford.edu/synop/), and the steady-state is obtained by 
iterating the MHD equations:

\begin{eqnarray}
&\frac{\partial \rho}{\partial t}+\nabla\cdot(\rho \mathbf{u})=0,&  \nonumber \\
&\rho \frac{\partial \mathbf{u}}{\partial t}+
\nabla\cdot\left( 
\rho\mathbf{u}\mathbf{u}+pI+\frac{B^2}{2\mu_0}I-\frac{\mathbf{B}\mathbf{B}}{\mu_0}
\right) = \rho\mathbf{g},& \nonumber \\
&\frac{\partial \mathbf{B}}{\partial t}+
\nabla\cdot(\mathbf{u}\mathbf{B}-\mathbf{B}\mathbf{u})=0, &  \\
&\frac{\partial }{\partial t}\left( 
\frac{1}{2}\rho u^2+\frac{1}{\gamma-1}p+\frac{B^2}{2\mu_0}
 \right)+ &
 \nonumber \\
&
\nabla\cdot\left( 
\frac{1}{2}\rho u^2\mathbf{u}+\frac{\gamma}{\gamma-1}p\mathbf{u}+
\frac{(\mathbf{B}\cdot\mathbf{B})\mathbf{u}-\mathbf{B}(\mathbf{B}\cdot\mathbf{u})}{\mu_0} 
\right)=\rho(\mathbf{g}\cdot\mathbf{u}), \nonumber &
\label{MHD}
\end{eqnarray}
where $\rho$, $\mathbf{u}$, $\mathbf{B}$, $p$, $\mathbf{g}$, and $\gamma$ are the mass density, velocity, magnetic field, 
pressure, gravity, and the polytropic index, respectively, until convergence is achieved.

We initiate the CME by superimposing an unstable, semi-circular flux rope based on the analytical model by \cite{titov99} 
on top of the ambient solution \citep{roussev03a}. The flux rope properties are matched to fit the observed properties of the 
source active region and its inversion line. The free energy is controlled by an additional toroidal field that produces 
the observed linear speed of the CME. The simulation here is based on a previous simulation done by \cite{cohen08a}, 
which includes full propagation of the CME to 1 AU and comparison with in-situ data. The CME initiation method described here 
has been used to study processes in the solar corona \citep[e.g.][]{lugaz07,manchester08}.

\subsection{Simulation Setup}

In order to follow the topology and evolution of the magnetic interaction between the CME and the ambient field, we design 
the grid with high resolution around the active region and along the line of CME propagation. The grid size around the active 
region is $10^{-3}R_\odot$, and the grid size along the propagation line is $10^{-2}R_\odot$. The grid resolution is coarser in 
other regions of the simulation domain, which are not of interest in the context of this work. The time step in the simulation 
is determined by the Alfv\'en speed and the grid resolution. The higher the magnetic field and smaller the grid size, the smaller 
the time step. In order to compensate for the extremely small time steps in active regions, where magnetic fields are very strong, 
we set the inner boundary of the simulation to be at a height of $0.06R_\odot$ above the surface, where the magnetic field is 
weaker.

\begin{figure}
\noindent\includegraphics[width=6.0in]{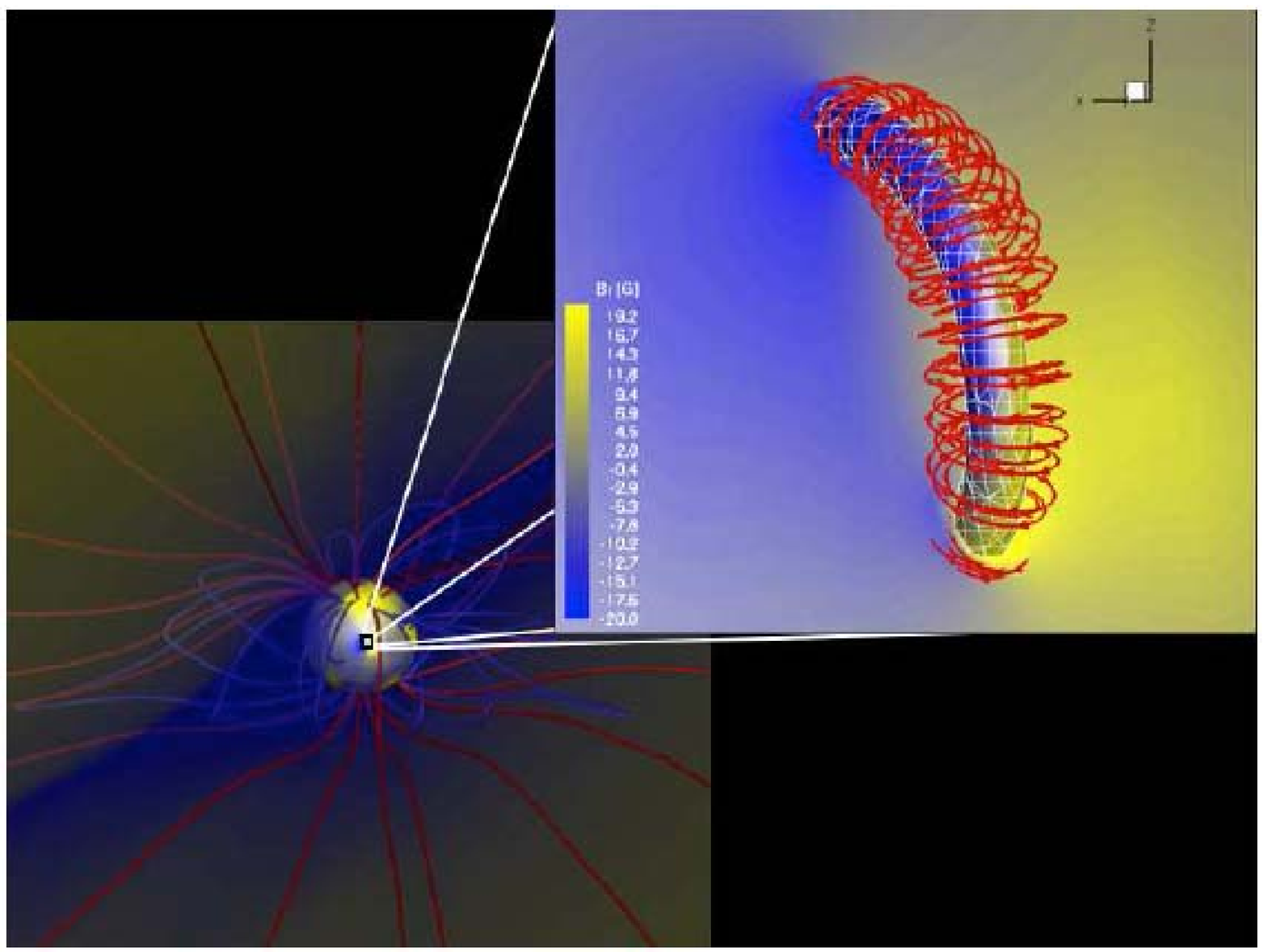}
\caption{
The left panel shows the large-scale structure of the steady-state corona. Selected field lines 
are shown in blue (closed field lines) and red (open field lines). Background color contours 
represent the solar wind radial speed (yellow-fast, blue-slow). The right panel shows the vicinity 
of the active region at the initial state of the simulation, where its approximated location on the 
Sun is marked by the black square. Color contours represent the magnitude of the radial field on a 
sphere at a height of $1.06R_\odot$, red streamlines represent three-dimensional magnetic field lines of 
the superimposed flux rope, and solid white lines mark the grid structure around the flux rope. 
The flux rope itself is represented by an iso surface of mass density, $\rho=2\cdot 10^{-14}\;g\;cm^{-3}$, 
which is greater than the surrounding density at the same height.}
\label{fig:f1}
\end{figure}

Figure~\ref{fig:f1} displays the initial stage of the simulation.  The left panel shows the large-scale structure of the steady-state corona, 
with selected closed field lines in blue and open field lines in red. The background color contours represent the solar wind radial 
speed, ranging from fast in yellow to slow in blue. The right panel shows the vicinity of the active region at the initial state of 
the simulation, where its approximated location on the Sun is marked by the black square. Color contours represent the magnitude of 
the radial field on a sphere at a height of $1.06R_\odot$, red streamlines represent three-dimensional magnetic field lines of the 
superimposed flux rope, and solid white lines mark the grid structure around the flux rope. The flux rope itself is represented by 
an iso surface of mass density, $\rho=10^{-14}\;g\;cm^{-3}$, which is greater than the surrounding density at the same height.

\subsection{The Concept of Numerical Reconnection in the Global Model for the Solar Corona}

The model solves the set of ideal MHD equations, so no physical resistivity is introduced in the equations. Therefore, the 
rate of {\it numerical} magnetic reconnection in the simulation is, in principle, controlled by numerical diffusion term, 
which is designed to stabilize the numerical solution and is not directly related to the physical resistivity of the system. 
The value of the numerical diffusion is proportional to the size of the grid cell and the time step via the ratio  
$\Delta x^2/\Delta t$. In the case of the numerical resistivity in the induction equation (where it can affect magnetic 
reconnections) this ratio is proportional to 
$\eta/\mu_0$, where $\mu_0$ is the permeability of free space, and $\eta$ is 
the electric resistivity. In the solar corona, the typical value for the Lundquist number, $S=\mu_0 Lv_A/\eta$ is $10^{12}-10^{14}$ 
\citep{BoydSanderson03}, 
where $L$ is the typical length scale, and $v_A$ is the Alfv\'en speed. The typical value of $\Delta x$ is a fraction of a solar 
radius and the time step can range between 
$0.1-10\;s$. Therefore, using the same typical values for $L$ and $v_A$ as above, we find that 
$1 < S_n = Lv_A \Delta t/\Delta^2 x\ll S$, where $S_n$ is the Lundquist number calculated using the numerical resistivity. This means 
that the model might result in over-reconnection of the magnetic field in regions where two opposite magnetic field lines are 
pushed towards each other. An example for such numerical behavior is the generation of ``U-shape'' detached field lines 
around the heliospheric current sheet. This issue is resolved by implementing the Roe solver in the numerical model, which is 
a more precise, less diffusive numerical scheme, and is equivalent to the addition of one more level of grid refinement 
(regardless of the actual smallest grid size). The Roe solver and its implementation to the MHD model are discussed in details 
in \cite{Sokolov08}.  

When discussing the magnetic reconnection of a CME with the coronal ambient field, one should keep in mind that this reconnection has 
a global, macroscopic sense, and that the time scale for such reconnection (which can be of the same order as the Alfv\'enic 
time scale) depends on the CME size and speed. In the simulation presented here, we try to capture this global interaction 
between the newly injected magnetic flux carried by the CME and the pre-existing field. This concept is different from 
the microscopic description of magnetic reconnection that obviously needs a better treatment than a global MHD model. In this simulation, 
we focus our study on the re-distribution of the global field as the new flux is introduced into the system, and on how the interaction between 
the fields affects the propagation of the CME through the corona. We compare our results with observations of large-scale signatures for such 
interaction between the fields in the corona. We emphasis that any further mentioning in the text for the term ``magnetic reconnection'' refers 
to a numerical reconnection in the simulation. 
 
We simulated the first 20 hours of propagation using the PLEIADES cluster at NASA's NAS center. The CME front left the simulation 
domain after approximately 14 hours. 
 

\section{Results}
\label{sec:Results}

We separate the analysis of the results into two parts. First, we validate the timing and the overall structure of the CME 
with some observations, and, second, we analyze the results assuming the overall interaction is satisfactorily captured by 
the simulation. Figure 2 compares the observed and simulated coronal dimmings and the observed and simulated coronal wave front 
associated with the density changes due to the CME expansion and propagation through the corona [after Cohen et al., 2009]. 
The four rows display comparisons for the early stages of the simulation, 10, 30, 40, and 60 minutes after the eruption onset, 
from top to bottom.  The first column shows SOHO EIT $195\mathring{A}$ base-difference images, constructed with data from
http://sohowww.nascom.nasa.gov. The second column shows the corresponding synthetic EIT $195\mathring{A}$ base-difference 
images produced by the simulation. These images are the line of sight integral: $\int n^2_eR(n_e,T)dl$, where $n_e$ is 
the electron number density and $R$ is the response function for the EIT $195\mathring{A}$ filter based on spectral 
synthesis from the CHIANTI code \citep{landi02}. 
The complete description of the synthetic EIT images in SWMF can be found in \cite{downs10}. The third 
column shows base-difference images of the simulated mass density on a sphere at height of $r=1.1R_\odot$. 
The differences are normalized to the pre-eruption mass density distribution in a similar manner as in \cite{cohen09}. 
The last column shows a display that is similar to the third column but with the addition of selected magnetic field lines. 
Those in the core flux rope are marked in red, those with one footpoint rooted in the original source region are marked in blue, 
and overlying field lines with both footpoints located far from the source region are marked in yellow. The approximate location 
of the leading edge of the coronal wave disturbance is marked by a white circle in the first and the third columns.  

\begin{figure}
\noindent\includegraphics[width=6.in]{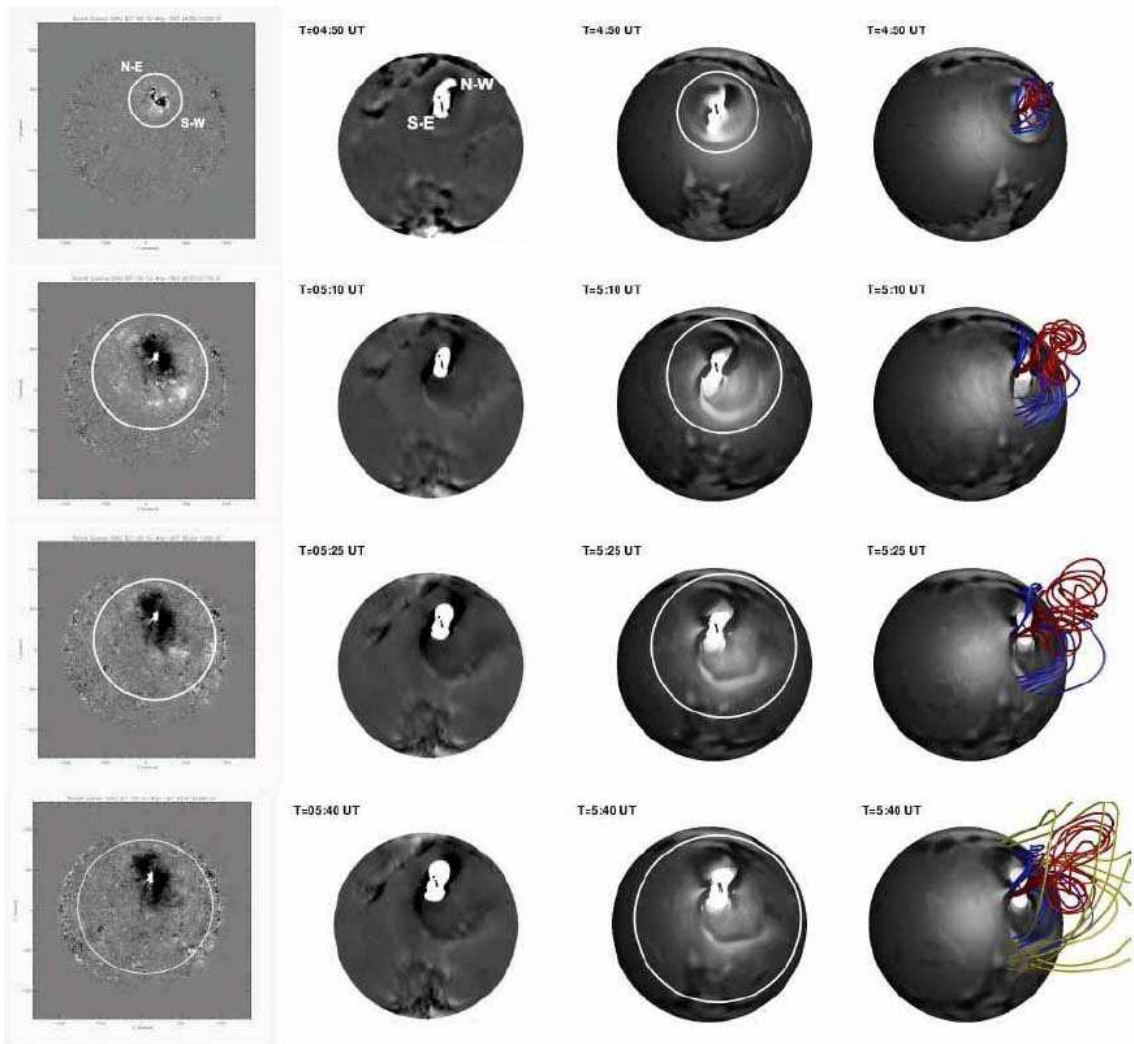}
\caption{
Comparison of observed and simulated coronal dimmings and the coronal wave for the early stage of 
the simulation: 10, 30, 40, and 60 minutes after the eruption onset (top to bottom). Left column shows 
SOHO EIT $195\mathring{A}$ base-difference images, second column shows the corresponding synthetic EIT 
$195\mathring{A}$ base-difference images produced by the simulation. The third panel shows base-difference images of 
the simulated mass density on a sphere at height of $r = 1.1R_\odot$. The right column shows a similar 
display to the third panel, but with the addition of selected magnetic field lines of the core flux 
rope marked in red, selected field lines with one footpoint rooted in the original source region in 
blue, and selected overlying field lines with both footpoints located far from the source region in 
yellow. The white circle in the first three columns indicates the approximate location of the leading 
edge of the coronal wave disturbance.}
\label{fig:f2}
\end{figure}

Figure~\ref{fig:f2} shows that the strongest brightening in the synthetic EIT images (second column) lags behind the matching observed 
(first column) and simulated (third column) coronal wave fronts, and it remains there in a location comparable to the location 
of persistent brightenings observed in the base difference images in the first column \citep[see][]{attrill07a,delannee09}. 
Although the locations of the observed and simulated wave fronts match, the latter appears only as a weak density enhancement. 
On the other hand, the source region of the event appears bright both in the synthetic EIT images and in the density difference 
images.  A discrepancy between the observed and simulated patterns is the lack of any rotation of the coronal wave in the simulation. 
\cite{podladchikova05} and \cite{attrill07a} independently showed that in the EIT observations the coronal wave 
undergoes an overall counter-clockwise rotation. Another discrepancy concerns the coronal dimmings. The NW-SE orientation of the 
main pair of dimmings in the simulated images appears as the mirror reflection of those in the observations (NE-SW). Coronal dimmings 
located on either side of a post-eruption arcade are understood to be the footpoints of the flux rope \citep[e.g.][]{webb00}. 
A possible reason for these discrepancies is that the flux rope in the simulation is idealized and undergoes only a straightforward 
expansion.  Real flux ropes can be much more complicated, with twist and writhe that might lead to the observed rotation of the coronal 
wave and a different representation of the dimmings.

The global field topology as shown in the last column of Figure~\ref{fig:f2} is similar to the topology extrapolated using a potential field 
method \cite{delannee09}(her Figure 2). This is not surprising, since the ambient magnetic field lines in the low corona are almost 
potential, even in the MHD solution. The main differences between the MHD solution and the potential field appear in the field lines 
that are stretched by the solar wind in the high corona and in the field lines of the superimposed flux rope, which are twisted. 
Unlike the potential field method, the time-dependent, MHD simulation enables us to follow the temporal evolution of the magnetic 
field and its dynamic response to the CME expansion.

\begin{figure}
\noindent\includegraphics[width=4.in]{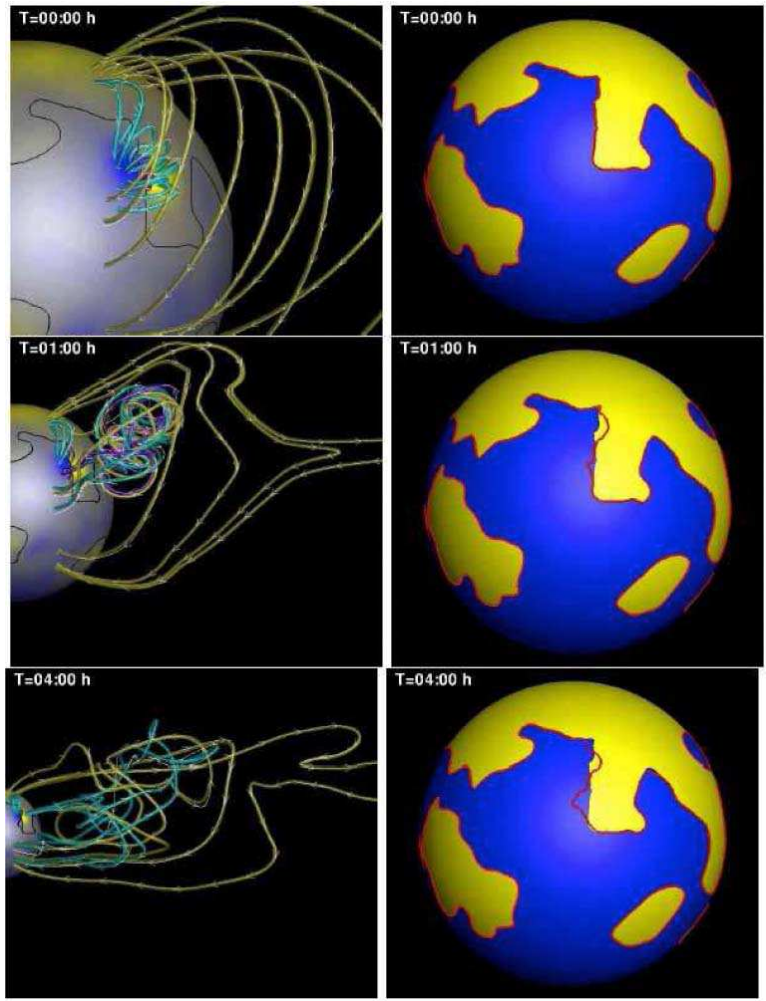}
\caption{
Evolution of the field topology at later time of the simulation than shown in Figure~\ref{fig:f2}. Left panel 
shows the photosphere colored with contours of $B_r$ intensity, along with selected magnetic field lines. 
Magenta field lines correspond to the core CME flux rope, with both footpoints rooted in the source 
region. Cyan lines mark overlying field with one footpoint located at the source region, and yellow 
lines mark field lines of the overlying helmet streamers and open field lines with footpoints located 
far from the source region. Right panel shows a sphere at a height of $r=1.06R_\odot$ colored with the sign 
of $B_r$ (positive-yellow, negative-blue) for the initial state of the simulation. The red line marks 
the inversion line for the particular time frame.}
\label{fig:f3}
\end{figure}

Figure~\ref{fig:f3} shows snapshots of the evolution of the field topology from the start of the simulation (row 1) to one (row 2) and four (row 3) 
hours later. The left column shows the photosphere colored with contours of $B_r$ intensity along with selected magnetic field lines (which 
are not the same in all frames). Magenta marks field lines in the core CME flux rope, with both footpoints located at the source region. 
Cyan marks overlying field lines with one footpoint rooted in the source region, and yellow marks field lines of the overlying helmet 
streamers and open field lines with footpoints located far from the source region. Row 2 shows that after one hour, some legs of the 
three-dimensional CME (represented by the highly twisted field lines in cyan, magenta or yellow) have expanded beyond the original active 
region so that there are more cyan field lines and fewer magenta ones. After four hours, row 3 shows that many field lines of the CME are 
located far from the source active region, though we emphasize that some still remain rooted at the original source AR. While the CME 
dynamic expansion pushes the surrounding field lines to the sides, the only way the footpoint of the CME can migrate away from the source 
AR is through reconnection with surrounding field lines as the CME is expanding. By the time the CME reaches the height of the large helmet 
streamers and the open flux that have opposite polarity to that of the CME, the CME footpoint expansion stops and only the traditional 
volumetric expansion of the CME continues. The overall field evolution at this point in the simulation is roughly consistent with the 
idealized scenario suggested by the breakout model, probably due to the similar field topologies of the idealized and real cases. 

The right column of Figure~\ref{fig:f3} shows a sphere at height $r = 1.06R_\odot$ colored with the sign of $B_r$ for the initial state of the simulation, 
where positive is yellow and negative is blue. Overlain on this initial polarity pattern is the time-evolving polarity inversion line in 
red. Rows 2 and 3 show that the inversion line changes as the CME propagates and interacts with the surrounding field. This change in 
the distribution of positive-negative flux reflects the exchange of closed flux between the CME and the surrounding field.  

In order to validate the simulated CME flux rope as it propagates into interplanetary space, we extract the simulated plasma parameters 
across the flux rope (upstream to downstream) from the frame of t = 10h after the eruption onset. We cannot directly compare our 
simulation result with 1 AU in-situ data, since the simulated CME only reaches $\sim 23-24R_\odot$, and the coordinate systems of the two data 
sets are different (Heliocentric Carrington system in the simulation vs. Geocentric system for in-situ data). However, we can compare 
the flux rope orientation and chirality through the local $B_z$ component and assume these should be conserved from the simulated location 
of the CME to 1 AU. 

\begin{figure}
\noindent\includegraphics[width=6.in]{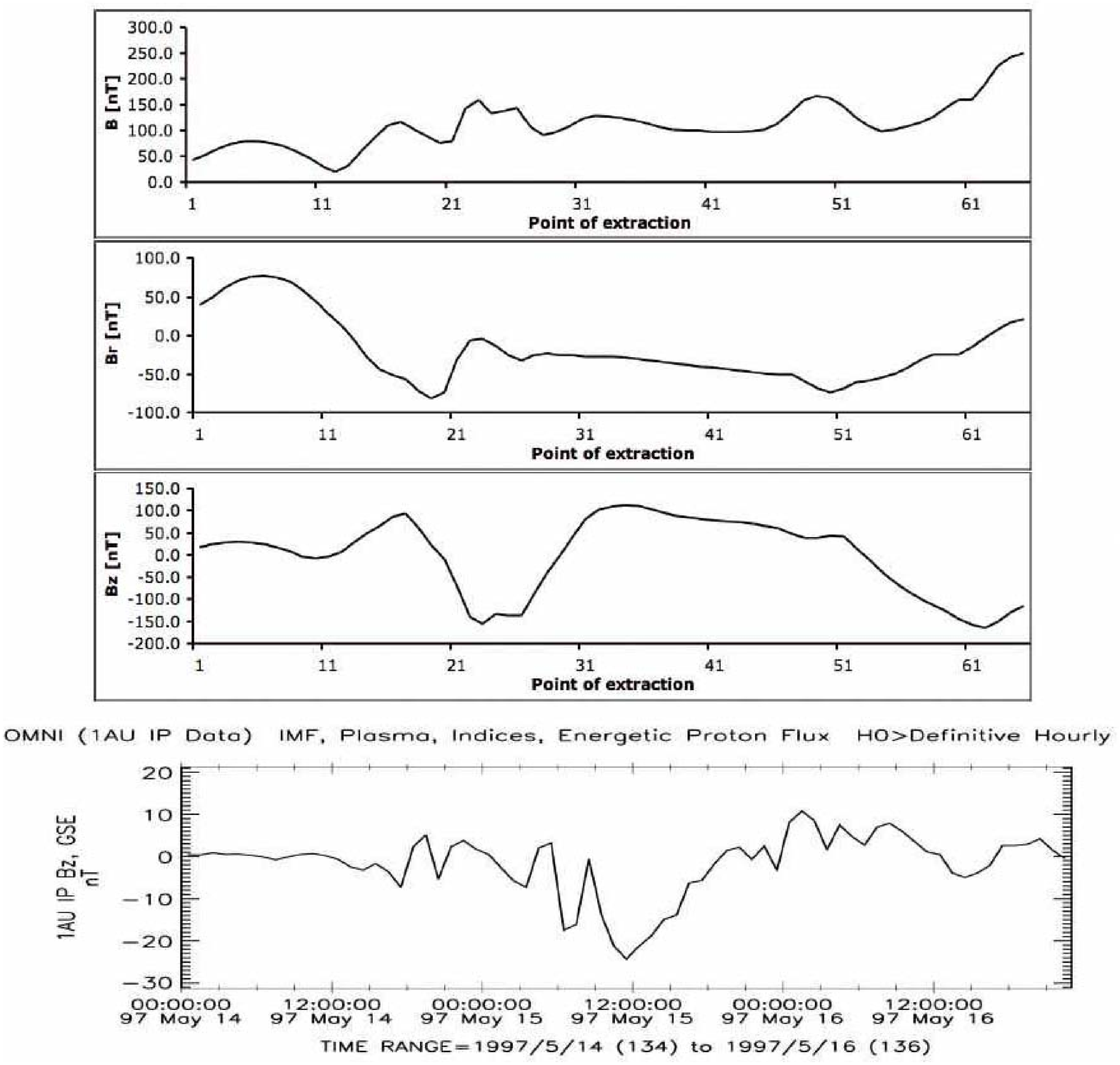}
\caption{
First three panels show upstream to downstream line extraction of $B$, $B_r$, and $B_z$, respectively, 
in the simulation frame of reference. The fourth panel shows the GSE $B_z$ observed by the Wind spacecraft 
between May 14-16.}
\label{fig:f4}
\end{figure}

The first three panels of Figure~\ref{fig:f4} show a line extraction of $B$, $B_r$, and $B_z$ along constant latitude and longitude from upstream 
(un-shocked plasma) to downstream (shocked plasma) in the simulation frame of reference. The fourth panel shows the GSE $B_z$ component 
observed by the Wind spacecraft during May 14-16 (data taken from http://cdaweb.gsfc.nasa.gov/). It can be seen that the field rotation, 
indicated by the change of sign in $B_z$, is present in the simulation as in the observations (compare panels three and four). However, we 
note that the magnitude of the simulated $B_z$ is greater than in the observations. This may be attributed to the simulation data being 
extracted closer to the Sun than the 1 AU observations. Here we do not attempt to match the result of the model with this observation, 
which has been taken at different location in the heliosphere and in different coordinate system, only to demonstrate that the field 
rotation seen in the data also appears in the simulation.  

\begin{figure}
\noindent\includegraphics[width=6.in]{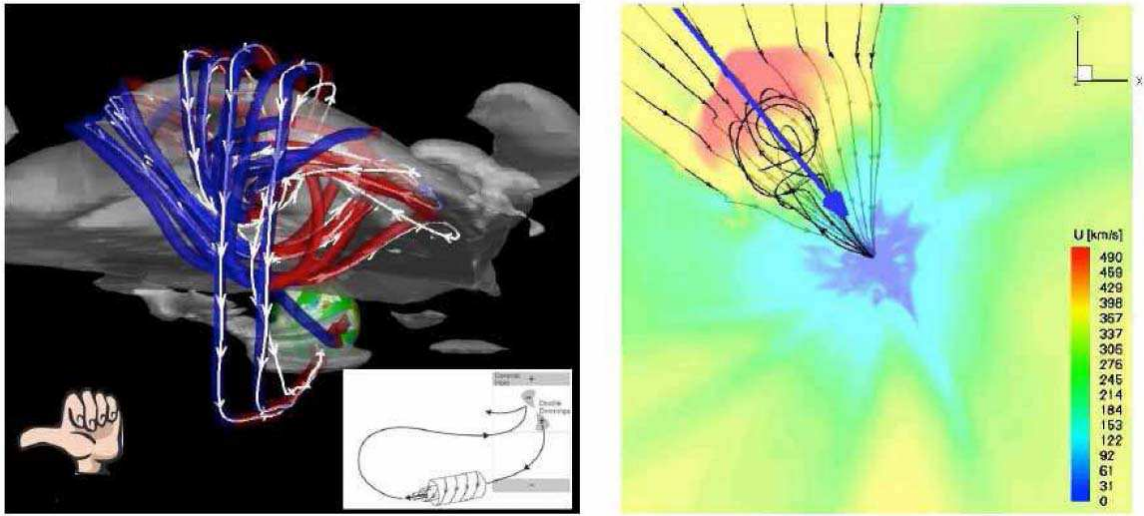}
\caption{
The field topology at t = 10h. Left panel shows the photosphere as the inner sphere colored with the 
radial field line contours. The approximate CME front is shown by the white shade, which indicates an 
iso-surface of density ratio of 4 between the t = 10h frame and the initial state. Selected field lines 
show the front of the CME and the field topology. Blue color indicates negative value of $B_z$, red color 
indicates positive value of $B_z$, and the left hand cartoon represents the flux rope chirality. Inset shows 
the observed interplanetary magnetic cloud chirality and orientation (from \cite{crooker02}. 
Right panel shows a view on the equatorial plane colored with contours of the solar wind speed together with 
selected three-dimensional field lines. The blue arrow marks the upstream-downstream line of extraction shown in 
Figure~\ref{fig:f4}}
\label{fig:f5}
\end{figure}

A visualization of the field topology (also showing the line of extraction) is displayed in Figure~\ref{fig:f5}. The left panel shows the photosphere 
as the inner sphere colored by radial field strength. The approximate CME front is shown by the gray shading, which outlines an iso-surface 
of density ratio equal to 4 between the t = 10h frame and the initial state. Selected field lines illustrate the field topology. Blue and red 
indicate negative and positive values of $B_z$, respectively, and the cartoon of a hand indicates the 
left-handed chirality of the flux rope, where the thumb points in the direction of the axial field and the fingers curl in the direction of 
the coils. These are in agreement with the line plots of $B_z$ in Figure~\ref{fig:f4}, representing a transition from a downward field (negative $B_z$) at the 
front of the CME to an upward field (positive $B_z$) at the back. The signature is a smooth rotation of the field direction, indicating the passage 
across the core of a flux rope. This is one of the most distinctive signatures of a magnetic cloud \citep{burlaga81}. The CME topology in 
the left panel of Figure~\ref{fig:f5} is also consistent with the observed interplanetary magnetic cloud shown in the white inset in the lower right corner 
(from \cite{crooker08}). The right panel of Figure~\ref{fig:f5} shows a view looking down on the ecliptic plane, which is colored with contours of the 
solar wind speed. Selected field lines are shown in black, and the blue arrow indicates the line of extraction used for the plots in Figure~\ref{fig:f4}.

\begin{figure}
\noindent\includegraphics[width=6.in]{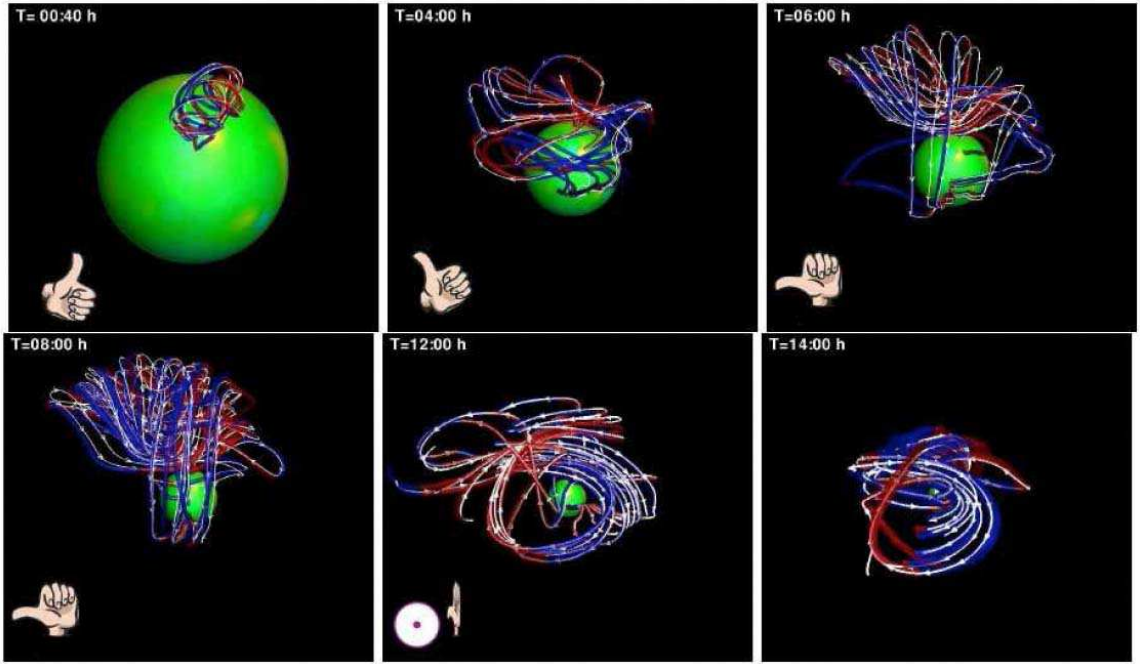}
\caption{
The orientation of the flux rope at different stages of the simulation. Display is similar to the left panel 
of Figure~\ref{fig:f5}. Blue (red) indicates negative (positive) $B_z$. The orientation of the flux rope axis is initially 
$\sim$N-S (cf. Figure~\ref{fig:f1}). By 00:40 h, the orientation of the axis is NW-SE. In the frame at 04:00 h, the 
orientation is hard to establish, but by 06:00 h, the axis of the flux rope lies $\sim$E-W. The last two 
frames at 12:00 h and 14:00 h are taken when most of the CME has left the simulation box.}
\label{fig:f6}
\end{figure}

Figure~\ref{fig:f6} shows the orientation of the flux rope at different stages of the simulation, where the display is similar to the left panel of Figure~\ref{fig:f5}. 
It can be seen that the rotation is well established about 6 hours after the eruption onset and that, after this, the CME maintains its orientation 
(it can still rotate around the z-axis). This result is also consistent with the overall behavior of the flux rope rotation described by 
\cite{lynch09}, who used the breakout model in a topologically idealized numerical simulation to study the flux rope rotation. In their model, 
a rotation of $\sim 40-50$ degrees is established during the first $\sim$ 20 minutes of the flare onset (co-temporal with an increase in the magnitude of the 
poloidal magnetic field). They ascribe the rotation primarily to the Lorentz force and overall torque from the sheared core magnetic field and 
related tension forces and conclude that the kink instability \citep[e.g.][]{kliem04}, \citep[see also][]{green07} is not responsible. We 
cannot study the possible effects of the kink instability in our simulation because the Titov-Demoulin flux rope is idealized and no kink-unstable twist is introduced. 
Our simulation results seem to imply that the flux rope rotation is strongly influenced by interaction with the surrounding field, since the 
rotation is established at the early stage of the CME propagation, between 04:00 and 06:00 h (Figure~\ref{fig:f6}).

\begin{figure}
\noindent\includegraphics[width=6.in]{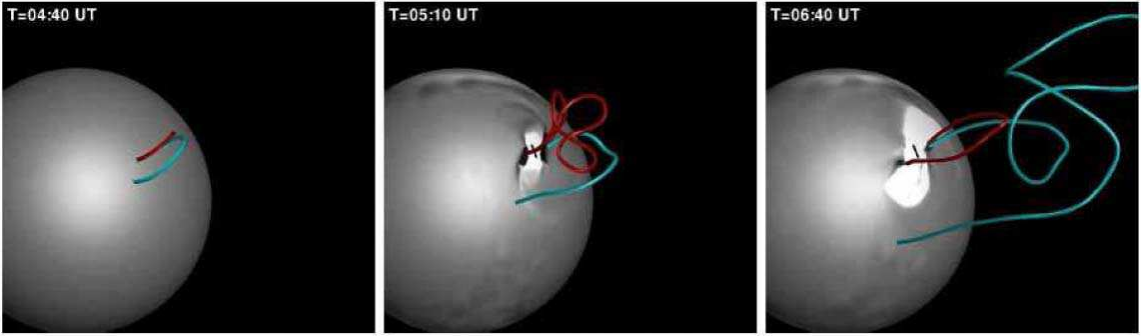}
\caption{
Displacement of the CME footpoints via stepping reconnection with neighboring magnetic field. 
The figure shows selected field lines and their evolution in time. Density base differences at a 
height of $r = 1.1R_\odot$ are shown in grey (as in Figure~\ref{fig:f2}). In the left and middle panels, the red line 
indicates the flux rope and has both footpoints rooted in the source AR. The cyan field line has 
only one footpoint rooted in the source AR, and represents a neighboring field line. In all frames, 
twisted field lines are associated with the CME. The CME initially expands with its flux rope footpoints 
(red) located near to the active region (middle panel). Later on (right panel), reconnection has 
occurred between the expanding flux rope (red), and the neighboring field line (cyan). This mechanism 
migrates the footpoints of the field lines associated with the CME away from the source AR and 
facilitates a lateral expansion of the CME. The CME footprint thus expands to a large extent in the 
low corona.}
\label{fig:f7}
\end{figure}

In Figure~\ref{fig:f7}, we display selected field 
lines and show their evolution in time in order to illustrate the stepping reconnection between the CME and adjacent loops, as described in 
Section~\ref{sec:Intro}. Unlike the field lines in Figure~\ref{fig:f3}, here the same field lines are displayed in the different time frames. 
The sphere represents the 
photosphere colored with the density base differences as in the right panel of Figure~\ref{fig:f2}. In the left and middle panels of Figure~\ref{fig:f7}, 
the red line 
indicates the flux rope and has both footpoints rooted in the source AR. The cyan field line represents a neighboring field line with only one 
footpoint rooted in the source AR. The CME initially expands with its flux rope footpoints (red) located in the active region (middle panel). 
Later on (right panel), reconnection has occurred between the expanding flux rope (red), and the neighboring field line (cyan). This reconnection 
transfers part of the original twisted flux rope field to the neighboring loop. The expanded flux rope now has only one footpoint in the source 
active region. The other has been ``stepped out'' and is now rooted in the quiet Sun at a considerable distance from the active region. The 
results displayed in Figure~\ref{fig:f7} give an example of how one field line in a CME expands laterally as a result of magnetic reconnection between the 
core flux rope and surrounding magnetic field. This mechanism migrates many of the footpoints of the field lines associated with the CME away 
from the source AR and thus facilitates a global lateral expansion of the CME in the low corona \citep{vandriel-gesztelyi08}.


\section{Discussion}
\label{sec:Discussion}

The results of our numerical simulation show that in the low corona the CME interacts with the surrounding field, which is composed of magnetic 
loops with different sizes, orientations, and mixed polarity. The path of CME expansion in this region is determined by the easiest way the CME 
can get through the surrounding field structure. This path depends on where the magnetic tension is smallest (i.e., in larger overlying loops) 
and where the field topology allows the CME to reconnect with surrounding field lines. The latter leads to a lateral expansion of the CME via 
migration of the CME footpoints. In addition, the simulation results show that the left-handed flux rope rotates counterclockwise by 90 degrees 
by the 
time the CME has expanded into interplanetary space so that the original $\sim$ N-S orientation of the flux rope (see Figure~\ref{fig:f1}), 
is now aligned $\sim$ E-W. 
This is consistent with the orientation of the magnetic cloud in the ecliptic plane deduced from observations \citep{webb00} and from 
modeling \citep{attrill06}. Both the CME lateral expansion and rotation wane when the CME starts to interact with the more organized 
magnetic flux of the large-scale helmet streamers and open flux. 

It is reasonable to believe that since both the CME rotation and expansion 
in our simulation are the result of the interaction between the expanding flux rope and the ambient flux via magnetic reconnection, the sense 
of the rotation is related to the chirality of the flux rope as well as the orientation of the ambient field, as suggested by \cite{lynch05}. 
We further suggest that the final orientation of the flux rope represents the ``relaxed'' state of the reconnection process between the core 
flux rope and the surrounding coronal field. 

Overall, our simulation produces results that are consistent with the breakout model, primarily because the topology of the ambient field 
and the core flux rope for the May 12, 1997, event is similar to the idealized field topology in the breakout model. One should keep in 
mind, however, that this might not be the case during solar maximum, when both the ambient field and the orientation of the core flux rope 
are unlikely to fulfill such idealized conditions. In addition, we stress that in the breakout model the field topology drives the eruption, 
while our simulation captures the post-eruption interaction between the CME and the ambient field.  

The result of the simulation suggests a scenario which enables us to develop a more complete picture of the May 12, 1997, eruption. 
Panels (1a) and (1b) of Figure 8 illustrate the scenario proposed earlier by \cite{crooker06a,attrill06}, and \cite{crooker08}, as well 
as \cite{owens07}. 
The CME, represented by the twisted line in panel (1a), expands and eventually undergoes interchange reconnection (ICX) with an open 
field line of the north polar coronal hole. As a result, the magnetic field in the CME that is rooted in the south dimming region is 
essentially an open field line (panel 1b), consistent with the open fields deduced from electron observations in the magnetic cloud at 1 AU. 
The accumulation of progressively larger closed loops between the northern dimming region  and the polar coronal hole has the effect of 
contracting the northern dimming region, which is well underway by 08:26 UT (see Figure 4 in \cite{attrill06}). The south dimming region 
remains near its maximum spatial extent for a longer time than the north dimming region, but by 15:00 UT, it has started to contract, as well. 
This contraction was not clearly addressed by \cite{crooker06a} and \cite{attrill06}. Why should the south dimming region should 
show a systematic contraction when the field lines rooted in that region should remain open (Figure 8, panel 1b)?

\begin{figure}
\noindent\includegraphics[width=7.in]{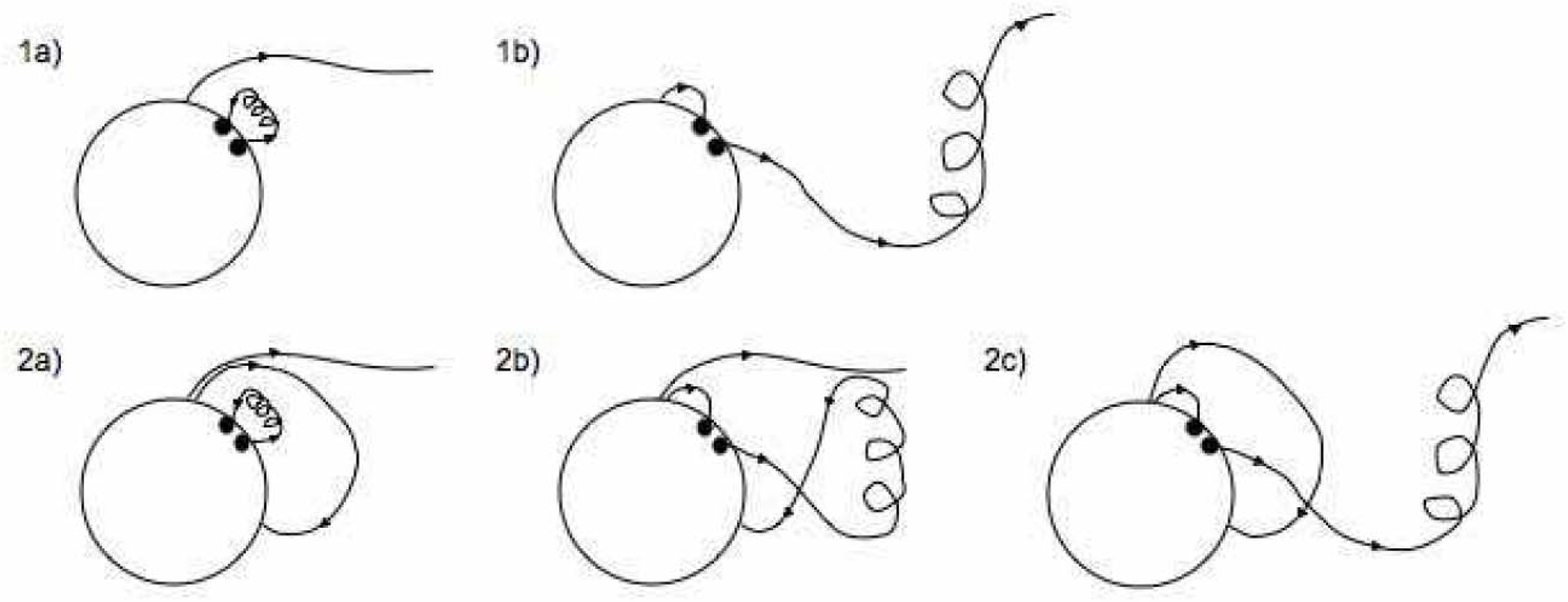}
\caption{
Panels (1a) and (1b) illustrate the scenario of the event as proposed by \cite{crooker06a,attrill06,crooker08}. 
The CME is represented by the twisted field line in panel (1a) and it undergoes ICX with an open field line of 
the north polar coronal hole. Panels (2a) and (2b) show the scenario based on the simulation results. The core flux rope 
first reconnects with an overlying closed field line, so closed loops (making up the north side lobe), form between 
the polar coronal hole and the north dimming region.  At the same time, the connectivity of the twisted closed loops that 
constitute the CME is changed, so that a significant portion of the CME is effectively displaced 
and expands out into the heliosphere. This extended twisted CME field can later reconnect with an 
open polar field line (panel 2c).}
\label{fig:f8}
\end{figure}

Our simulation suggests a slightly different scenario, that provides an answer to that question. In this scenario, 
illustrated in panels 2a and 2b of Figure~\ref{fig:f8}, the core flux rope first reconnects 
with greatly extended overlying closed field lines. As a result, the legs of the twisted closed loops that constitute 
the CME are displaced so that only the south dimming region remains connected to the greatly expanded twisted CME 
loops (panel 2b, Figure~\ref{fig:f8}). We note that a flux rope is made up of many complex twisted magnetic field lines. The 
outermost field lines reconnect with the surrounding magnetic field, but the innermost core field lines may not 
experience such reconnections. As a result, some part of the original flux rope may remain rooted in the source 
region throughout the eruption (not shown in Figure~\ref{fig:f8}). Indeed, such a picture is consistent with the deepest parts 
of coronal dimming regions remaining dimmed for several days \citep[e.g.][]{attrill08}, long after the main bulk 
of a CME has left the Sun. 

The contraction of the south dimming region in this May 12, 1997, event can now be explained as a result of successively 
larger closed loops being formed due to ongoing breakout-type reconnections, so contracting the south dimming in a 
similar manner to the north one, albeit at a later time due to the much larger scale of these closed loops.  During 
this process, one of the legs of the twisted field making up the CME is displaced out of the source AR. To our 
knowledge, this is the first time that a CME has been shown to comprise one of the side lobes created as a result 
of reconnection with the overlying field. Usually the central flux system in the breakout model erupts as the 
restraining overlying field is removed to the side lobes. Here the situation differs markedly from the textbook 
breakout scenario in several ways: (i) the reconnection occurs much closer to one footpoint of the overlying closed 
field than the other, introducing a marked asymmetry; and (ii) the overlying closed field is greatly extended so 
that upon reconnection, relatively long field lines constitute the side lobe and become dragged out into 
interplanetary space as the eruption proceeds.  These two factors mean that the north dimming region will start to 
contract, whilst the south region will remain extended for a longer time, as observed. 

What the new scenario in Figure~\ref{fig:f8} may struggle to explain is the unidirectional streaming of suprathermal electrons 
in the CME at 1~AU, indicating field lines connected to the Sun at only one end. Although the determination of 
magnetic connectivity from suprathermal electron data can be hindered by significant uncertainties 
\citep[e.g.][]{riley04}, in this case the signature of open fields throughout most of CME is quite clear. In contrast, 
our simulation results show that the extended twisted field identified as the CME is connected to the Sun at both 
ends, which topologically would be expected to produce counterstreaming electrons at 1~AU. Two possible explanations 
for this disagreement are: 1) Although not apparent in the simulation, the reconnection with overlying loops 
pictured in panel 1b of Figure~\ref{fig:f8} may proceed until the supply of closed loops gives way to the neighboring open polar 
field, in which case the scenario essentially reverts to the original one where interchange reconnection in the 
corona opens the loops (panel 2c, Figure~\ref{fig:f8}); 2) interchange reconnection opens the loops associated with the CME, 
but takes place gradually, in the solar wind, outside the simulation box, in transit to 1~AU. 

The first explanation suggests that the helmet streamer (forming the initial overlying closed magnetic field) is 
temporarily removed by the breakout process, and then later reforms through ICX with the open flux of the polar 
region. Even though we believe that the ICX scenario represented by panel 2c in Figure~\ref{fig:f8} must occur at some point, 
it is hard to capture this interaction in the current simulation (in contrast to the interaction with the closed flux). 
This may be due to the fact that the ICX interaction should occur closer to the outer boundary, where every field line that 
leaves the simulation domain is considered to be ``open''. We believe that the interaction of the CME flux with the 
open flux of the Sun can be better studied through idealized simulations, in which the isolation of such reconnection 
events can be better resolved, rather than in this complicated, realistic simulation.  In addition, we favor the first 
explanation over the second because reconnection in the solar wind does not act to balance the flux increase owing to 
the introduction of the CME loops \citep[e.g.][]{owens06}, nor does it transport open flux on the Sun in the 
course of the solar cycle \citep[e.g.][]{owens07}. At this stage, however, we leave the issue as an open question.


\section{Summary and Conclusions}
\label{sec:Conclusions}

Using numerical simulation, we study the evolution of a CME and its interaction with the coronal magnetic field 
based on the May 12, 1997, CME event. Our simulation provides the following results:

\begin{itemize}
\item The simulated density corresponding to a cut across the CME at $1.1R_\odot$ is comparable to the coronal wave 
bright front that appears in the observed EIT images. This is similar to results obtained for another 
event studied by \cite{cohen09}.
\item During the initial stage of CME expansion, the core flux rope reconnects with the surrounding (neighboring) 
field, which facilitates lateral expansion of the CME footprint in the low corona. Evidence for this is the 
displacement of the initial CME footpoints away from the source AR, via ``stepping reconnection''. 
\item The CME then reconnects with the oppositely orientated overlying magnetic field in the manner of the 
breakout model. This is due to the global field topology, which is essentially the same as the initial state of the 
breakout model. 
\item A significant component of the CME that expands into interplanetary space comprises one of the side lobes 
created mainly as a result of reconnection with the overlying field.
\item As the flux rope expands, it rotates counterclockwise by 90 degrees owing not to the kink instability but 
to reconnection with the surrounding fields.
\item The lateral expansion as well as the rotation of the flux rope due to interaction with the overlying magnetic 
field continue as long as the reconnections occur. They stop when the CME reaches the helmet streamers and the 
open flux, where the field of the CME matches the globally-organized field.
\item The orientation and left-handed chirality of the simulated flux rope are in agreement with the 
interplanetary flux rope observed at 1~AU.
\item The simulation shows no direct evidence of the anticipated interchange reconnection that would 
account for the electron observations indicating open fields in the CME at 1~AU.
\end{itemize}

From these results we conclude that reconnection between closed loops may play a major role in transforming 
a small-scale flux rope in an active region to a large-scale flux rope with widely-spaced footpoints and a 
new orientation before leaving the Sun as a CME. Therefore, this complex interaction should be taken into 
consideration in any attempt to predict the CME geo-effectiveness. We further conclude that simulations 
of CMEs that erupt into realistic background fields constructed from magnetograms afford unprecedented 
views of the complicated three-dimensional development of magnetic structure.
 

\begin{acknowledgments}

We thank for two referees for their useful comments and suggestions. 
OC and NUC are supported by SHINE through NSF grants ATM-0823592 and ATM-0553397. NAS was supported through 
the NASA LWS EMMREM project, grant no. NNX07AC14G. GDRA is supported by NASA grant NNX09AB11G-R. Simulation 
results were obtained using the Space Weather Modelling Framework, developed by the Center for Space 
Environment Modelling, at the University of Michigan with funding support from NASA ESS, NASA ESTO-CT, NSF KDI, 
and DoD MURI. SOHO is a project of international cooperation between ESA and NASA.
\end{acknowledgments}



\end{article}


\end{document}